## Kinetic-inductance Parametric Amplification in Superconducting Hybrid Magnonics

Pratap Kumar Pal[1,2], Moojune Song[1,3], Jinho Lim[2], Juliang Li[4], Serhii Kruhlov[4,5], Ralu Divan[6], Ulrich Welp[1], Clarence Chang[4], Kab-Jin Kim[3], Valentine Novosad[1,7,*], Axel Hoffmann[2,#], and Yi Li[1,$]

[1]Materials Science Division, Argonne National Laboratory, Lemont, IL 60439, USA

[2]Department of Materials Science and Engineering and Materials Research Laboratory, The Grainger College of Engineering, University of Illinois Urbana-Champaign, Urbana, IL, 61801, USA

[3]Korea Advanced Institute of Science and Technology: Daejeon, Daejeon, KR

[4]High Energy Physics Division, Argonne National Laboratory, Lemont, Illinois 60439, USA

[5]Department of Physics, Drexel University, Philadelphia, PA 19104, USA

[6]Center for Nanoscale Materials, Argonne National Laboratory, Argonne, IL, 60439, USA

[7]Institute of Multidisciplinary Research for Advanced Materials, Tohoku University, 2-1-1 Katahira, Aoba-ku, Sendai, 980-8577, Japan

[*]Email: novosad@anl.gov

[#]Email: axelh@illinois.edu

[$]Email: yili@anl.gov

### Abstract

Parametric amplification enables sensitive microwave detection and signal processing by amplifying weak signals with minimal added noise. Here, we demonstrate kinetic-inductance-mediated parametric amplification in a hybrid magnonic system comprising a nonlinear NbN nanowire resonator strongly coupled to a yttrium iron garnet (YIG) sphere. The resonator exhibits four-wave mixing and nonlinear frequency pulling, enabling parametric amplification both away from and within the magnon–photon strong-coupling regime. In the magnon-detuned regime, we observe 7 dB gain, while coupling to the YIG magnon shifts the parametric-amplification response by ~20 MHz and yields 3.4 dB gain in the hybrid mode. Thus, magnon–photon coupling provides a route to magnetically tune the spectral position of kinetic-inductance parametric amplification. At higher pump powers, the system transitions to a self-oscillatory regime featuring comb-like sidebands and probe-activated nonlinear mode mixing. These results establish hybrid magnonics as a platform for tunable nonlinear microwave amplification and dynamics.

## Introduction

Parametric amplification is a cornerstone of quantum microwave technology, enabling the amplification of extremely weak signals with minimum noise allowed by quantum mechanics[1]. Such amplifiers exploit a strong pump tone to drive a nonlinear element, coherently redistributing energy into a weak signal and its idler, thereby achieving high gain without introducing thermal noise from a DC current bias as arises in conventional transistor-based amplifiers [1–4]. This capability is crucial for modern quantum measurements: for example, superconducting qubits and spin-resonance experiments generate only a few microwave photons of signal, thus low-noise amplification is required to attain high signal-to-noise readout.

In the standard technology, Josephson-junction parametric amplifiers (JPAs) and traveling-wave parametric amplifiers[5,6] are routinely used in circuit QED for fast qubit readout while maintaining high fidelity[7]. Recently, a new technique of parametric amplification has been investigated using the intrinsic kinetic inductance in superconductors as the source of Kerr nonlinearity[8–10], offering high magnetic field compatibility for experiments involving large static fields. For example, spin qubit systems[11], quantum dot qubit devices[12], topological qubits[13], all require strong magnetic fields in the sample region, yet also need quantum-limited readout.

In this work, we explore the operation of a kinetic inductance parametric amplifier (KIPA) that is strongly coupled to magnetic excitations, or magnons[14]. The field of cavity magnonics [15–20] has led to many interesting physical demonstrations, particularly magnon-qubit entanglement [21–25]. This enables single-magnon quantum operations and provides new approaches to quantum magnonics, such as on-chip microwave isolators [26,27], microwave-to-optic quantum transduction [28–31], and magnon quantum sensing for dark matter [32–34]. When a KIPA is coupled to magnons, the hybrid modes inherit the frequency tunability of magnons, allowing for adjustable amplification bandwidth by an external magnetic field. Using 4-wave mixing conditions, we obtain a parametric amplification gain of 7 dB for the decoupled resonator, and a gain of 3.4 dB for the resonator when it is strongly coupled to the magnon mode. An amplification band shift of 20 MHz is induced by the magnon-photon mode hybridization. Our results provide a new avenue towards frequency-tunable parametric amplifiers along with magnetic field compatibility, which may find future applications in quantum information science and quantum magnonics.

## Results and Discussion

One major challenge of coupling a KIPA with magnon mode is the trade-off between kinetic inductance and geometric inductance in the resonator design, where the former is needed for increasing the resonator nonlinearity and the latter is needed for dipolar coupling to magnons. Our design is based on linking a circular antenna to the current antinode of a NbN ring resonator to achieve coupling to magnons [35]. In the current layout we implemented a much narrower signal line with width of 500 nm and a thinner NbN film of 35 nm for enhancing the kinetic inductance. A single-crystal YIG sphere with a diameter of 250 μm is mounted at the center of the circular antenna as described in our prior work[35]. Figure 1(a) illustrates the hybrid magnon–resonator device and its basic characterization. The superconducting nanowire resonator is placed in-plane and is coupled to a YIG sphere mounted on the same circuit, with one coplanar waveguide outlet coupling port for microwave signal readout. The resonator is then coupled to a vertical circuit antenna for implementing transmission measurements and pump injection. For parametric amplification measurements, a pump tone ($f_{pump}$) and a probe tone ($f_{probe}$) are mixed and injected to the vertical circular antenna for interacting with the in-plane superconducting nanowire resonator, with the microwave signals from the superconducting nanowire resonator measured by a spectrum analyser (SA). The design was originally planned for pump injection through the magnon mode, but experimentally we find that the crosstalk between the vertical antenna and the in-plane superconducting nanowire resonator is sufficient for implementing parametric amplification even when the magnon mode is detuned away from the resonator mode. For the convenience of understanding the measurements, we also show the linear power transmission spectrum measured by a vector network analyser (VNA) from the vertical antenna to the in-plane resonator. Figure 1(b) shows the VNA spectra as a function of external biasing field which is applied in the resonator plane for minimizing superconducting vortex generation. A magnon-photon coupling strength of up to $g/2\pi$ = 8.8 MHz is obtained as shown in the field-swept mode anticrossing between the Kittel magnon mode and the resonator mode. The large suppression of coupling strength compared with our previous report of 88 MHz [35] is due to the narrow signal line and the associated large kinetic inductance and more pronounced ground shielding, but this is necessary for the KIPA to function.

**Parametric amplification in nanowire superconducting resonator**. We first show the parametric amplification measurements at H=0 Oe, where the magnon mode is detuned and decoupled from the superconducting nanowire resonator. We use the four-wave mixing (4WM) pumping approach for parametric amplification as illustrated in Figure 2(a). The pump tone ($f_{pump}$) close to the resonator frequency is applied together with the weak probe tone ($f_{probe}$) positioned in its vicinity. Owing to the Kerr-type nonlinearity arising from the kinetic inductance of the superconducting nanowire, the driven resonator supports a four-wave mixing process in which two pump photons are converted into a probe photon and an idler photon satisfying the energy-conservation condition ($f_{probe} + f_{idler} = 2f_{pump}$). Under the appropriate phase-matching and resonance conditions, energy is transferred from the pump to the probe and idler frequencies, resulting in amplification of the probe signal. Figures 2(c–f) present two-dimensional ($f_{probe}$ vs. $f_{SA}$) heatmaps of the spectrum analyser (SA) data at four different $f_{pump}$ values under the same pump condition ($P_{pump}$ = -22 dBm), and with varying $f_{probe}$. The diagonal spectrum shows the dependence on $f_{probe}$, and the horizontal red line shows the dependence on $f_{pump}$. When $f_{pump}$ is off-resonance (cases 1 and 4), the probe response shows a Lorentzian resonance peak centered at 4.3413 GHz, and the peak amplitude evolution of $f_{probe}$ reproduces the intrinsic VNA transmission spectrum of the resonator without pump. When $f_{pump}$ is within the resonance bandwidth of the superconducting resonator and approaches the resonance frequency from the lower-frequency side, a clear frequency pulling is observed (case 2, $f_{pump}$ = 4.3403 GHz), i.e., the microwave transmission maximum for the $f_{probe}$ tone shifts towards lower frequency by around 1.5 MHz as compared with case 1 in Fig. 2(c) [also shown in Fig. 2(b) as discussed later]. This frequency pulling is accompanied with an X-shape spectral pattern, indicating the emergence of the idler mode and the associated 4WM process.

Figure 2(b) plots the peak amplitude of the SA response at each probe frequency: for a given $f_{probe}$, we record the SA spectrum and extract the peak power amplitude at $f_{probe}$; repeating this for successive values of $f_{probe}$ yields the curves shown in Fig. 2(b) for the four $f_{probe}$ conditions corresponding to Fig. 2(c–f). By comparing the frequency pulling cases (cases 2 and 3) with the off-resonance cases (cases 1 and 4), a clear enhancement of the maximal power transmission peak value is observed, suggesting an amplification of the transmitted microwave at $f_{probe}$. The gain, defined as the peak difference of the transmitted probe signal between the on-resonance and off-resonance cases, is maximized to 7 dB at $f_{pump}$ = 4.3403 GHz, then decreases to 3 dB as $f_{pump}$ moves up to $f_{pump}$ = 4.3405 GHz, and eventually goes to zero as $f_{pump}$ moves outside the resonance regime. The gain is independent of $P_{probe}$ in our measurements. The simultaneous appearance of frequency pulling, X-shape spectral pattern, and finite gain demonstrate parametric amplification due to 4WM, and reinforces that energy is transferred from the pump into both the probe and idler mode. We also point out that the gain is associated with $P_{pump}$. At lower $P_{pump}$, both the frequency pulling and the gain will decrease. When $P_{pump}$ is increased above -22 dBm, the frequency pulling will become larger but the gain no longer increases because $P_{pump}$ is approaching the spontaneous instability threshold and the 4WM process cannot pump more energy from $f_{pump}$ into $f_{probe}$. The spontaneous instability regime will be discussed later.

Figure 2(g) and (h) show how 4-wave mixing and its associated parametric amplification is triggered. In both cases, the probe frequency is set to be $f_{probe}$= 4.3399 GHz and the pump frequency is shifted from $f_{pump}$= 4.3401 GHz in case 1 to $f_{probe}$= 4.3403 GHz in case 2, with the SA lineshapes marked by white dashed lines in Fig. 2(c) and (e), respectively. For case 1, with the pump frequency set just outside the nonlinear frequency pulling range, the spectrum contains the pump and probe tones without the appearance of an additional mixing product, indicating that the four-wave mixing does not happen. By slight shifting of the pump frequency in case 2 which is now within the frequency pulling range, an additional idler tone emerges at the frequency expected from the four-wave-mixing relation ($f_{idler} = 2f_{pump} - f_{probe}$), while the probe response is simultaneously enhanced. The direct comparison between Figures 2(g) and (h) therefore demonstrates the strong frequency dependence of the nonlinear mixing process: a small change in pump frequency is sufficient to switch the system from the linear state to the nonlinear four-wave-mixing regime.

As a reference for understanding the nonlinear frequency pulling, we also performed a two-tone VNA measurement, in which the VNA signal is used as the probe tone and combined with another pump tone to be injected into the perpendicular circular antenna, with the transmitted signal from the in-plane superconducting resonator measured by the VNA. Figure 2(i) shows the results as a function of $f_{pump}$ at H = 0 kOe and $P_{pump}$=−22 dBm. The diagonal of the 2D color plot corresponds to $f_{pump} = f_{probe}$. When $f_{pump}$ reaches the threshold at $f_{pump}$= 4.3403 GHz, as indicated in Fig. 2(e), frequency pulling of the resonant transmission peak takes place, and gradually recovers as $f_{pump}$ becomes higher, reproducing the trend shown in Fig. 2(c)-(f).

**Parametric amplification using hybrid magnon-photon modes**. The property of parametric amplification can be also be realized when the superconducting resonator mode is strongly coupled to the magnon mode. To demonstrate parametric amplification in the hybrid magnon-photon modes, we set the magnetic field such that the magnon is degenerate with the resonator mode. Figure 3(a) shows the VNA-measured frequency pulling of the hybrid magnonic modes under a pump tone of $P_{pump}$=-22 dBm with the magnon and photon modes at degeneracy at H = -1.3021 kOe. The two horizontal spectra featured at 4.321 GHz and 4.344 GHz correspond to the two hybrid magnon-photon modes. The difference in amplitude between the two modes, evident from the color contrast, arises from interference between the hybrid modes and background microwave crosstalk; see Fig. 1(b) for the pronounced contrast difference between the two hybrid mode branches at H=-1.300 kOe. As $f_{pump}$ intersects each hybrid magnon-photon mode which is indicated by the intersection of the diagonal line with a hybrid-mode branch, frequency pulling is observed for both the hybrid mode that is degenerate with $f_{pump}$ and the other hybrid mode, whose frequency differs from $f_{pump}$. This means the nonlinear property induced by the photon component of one hybrid mode can be transferred to the other hybrid mode even though the pump tone interacts directly with only one of the hybrid modes.

To search for evidence of 4WM which is essential for parametric amplification, we show the SA measurements in the four frequency pulling regimes in Figure 3(a), labelled as A, A', B, and B'. In each regime, we show the measurements at two $f_{pump}$ conditions, one with $f_{pump}$ outside the frequency pulling range and one with $f_{pump}$ inside the frequency pulling range, as similar to Figure 2(g) and (h). Figure 3(b-e) show the results. In addition to the SA measurements, we also extract the peak amplitude of the probe response as a function $f_{probe}$, and compare the evolution between two different $f_{pump}$ conditions, similar to Figure 2(b).

First, we focus on the regimes A and B where both $f_{pump}$ and $f_{probe}$ are coupled to the same magnon-photon hybrid mode. For the lower-frequency branch as shown in Figure 3(d), when the pump is located in the frequency pulling regime [$f_{pump}$=4.32 GHz, A(II)], clear X-shape mode mixing is observed by sweeping $f_{probe}$, indicating 4WM for the hybrid mode. As the pump frequency is moved slightly outside the frequency pulling regime [$f_{pump}$=4.318 GHz, A(I)], no mode mixing is observed. The 4WM process is accompanied with probe amplification, with the SA-measured probe amplitude shown in Figure 3(d), A(1). The maximum of the probe amplitude for the mixing condition is higher than the maximum of the probe amplitude for the no mixing condition by 3.4 dB, which is the signature of parametric amplification. We highlight that even strongly coupled with the magnon mode, the pump energy can still efficiently drive the photon mode component from the superconducting resonator into nonlinearity for probe gain. In the reciprocal regime of B, as shown in Figure 3(c) where the other hybrid magnon-photon mode is pumped, similar frequency pulling and 4WM can be observed, but there is no clear gain for the maximum of the $P_{probe}$ amplitude [Figure 3(c), B(3)]. This may be attributed to the complex role of background interference, which leads to a weaker lower-frequency hybrid magnon-photon branch as similarly shown in Figure 1(b). In one possible scenario frequency pulling lowers the maximal probe transmission frequency, which leads to even stronger destructive interference with the background crosstalk tone and hence suppresses the total power transmission coefficient.

For the regimes A' and B' where $f_{pump}$ and $f_{probe}$ are at different hybrid modes, SA measurements do not show signatures of 4WM. This is mainly because the expected idler mode is off the resonance and the 4WM process cannot be resonantly enhanced at $f_{idler}$. From the probe amplitude evolutions, the maximal power transmission frequency only shows a shift towards lower frequency as similarly shown by the VNA measurements in Figure 3(a), but the peak probe amplitude becomes lower after frequency pulling. To enable efficient 4WM and the associated parametric amplification, Davoudi et. al.[36] demonstrated a proof-of-concept amplifier based on two hybridized resonator modes, using off-resonant pumping at the center of the two hybrid mode and probing at either hybrid mode, so that both the probe and idler tones can sit inside hybrid modes. Similarly, Ref. [37] employed coupled resonator modes and three-wave mixing (3WM), where the pump frequency is set at the sum of the two hybrid modes for nondegenerate parametric amplification. In these approaches, hybridization between multiple resonator modes provides an additional degree of freedom for engineering the nonlinear interaction and spectrally separating the pump from the amplified signal.

Figure 3(f) shows the frequency tunability of parametric amplification by comparing the peak amplitudes of the probe responses between the hybrid magnon-photon mode [H = −1.3021 kOe, Figure 3(d)] and the uncoupled resonator mode [H = 0 kOe, Figure 2(b)]. The shaded areas show the regime where the power output of the pumped resonator is higher than the maximal power output level of the unpumped resonator mode, i.e. the regime with a true power amplification. Magnon-photon hybridization enables a large tuning range of the parametric amplification bandwidth center by 20 MHz, 15 times larger than the bandwidth itself (1.2 MHz). The gain decreases from 7 dB for the uncoupled resonator to 3.4 dB when the magnon mode strongly couples with the resonator mode. To quantitatively analyze the performance of the parametric amplifier, we calculate the gain-bandwidth product (GBP)[1,10,38,39] as GBP = (G-1)×B, where G is the maximal gain in the shaded area and B is the bandwidth of the shaded area in Fig. 3(f). Using the value of G=7.0 dB, B=1.2 MHz for the uncoupled resonator, and G=3.4 dB, B=1.6 MHz for the coupled resonator, we calculate GBP = 4.8 MHz for the uncoupled resonator and GBP = 1.9 MHz for the coupled resonator. This reduction is likely due to the hybridization of the superconducting nanowire resonator mode with the magnon mode, where the pump efficiency to the nanowire resonator is reduced in half. The operating frequency window of the parametric amplifier can be shifted and engineered without modifying the resonator geometry, highlighting an important advantage of hybrid magnonic platforms for frequency-agile microwave amplification.

**Self-oscillation at elevated pump power.**

According to theory[40], parametric amplification works for pump power approaching the self-oscillation threshold. To further confirm the pump power range, we extended the measurements to pump powers above −22 dBm. At sufficiently high pump power, the system undergoes a qualitative transition from the parametric-amplification regime to a self-oscillatory state. Figure 4(a) shows the spectrum analyzer measurements at $P_{pump}$=-16 dBm at the magnon-photon mode degeneracy condition as used in Figure 3 (H = -1.3021 kOe), with the VNA spectrum plotted in the background for comparison. Four different $f_{pump}$ are tested around the two hybrid magnon-photon modes. Self-oscillation is observed when $f_{pump}$ is slightly lower than the two hybrid mode peaks ($f_{pump}$=4.318 GHz and 4.343 GHz), corresponding to the frequency-pulling regime along with the highest gain in the parametric amplification regime. It is manifested as a series of cascaded peaks with even spacing, consistent with Hopf-like self-modulation or relaxation-oscillation behavior[41,42], and has been used as a frequency comb in many nonlinear dynamic systems [43–46]. The peaks have finite linewidths in the kHz range, different from the peak at $f_{pump}$ whose linewidth is equal to the microwave source of around 1 Hz. When $f_{pump}$ is at the center of the two hybrid modes ($f_{pump}$=4.324 GHz and 4.346 GHz), no self-oscillation is observed. This might be due to a lower effective gain at higher $f_{pump}$, such that self-oscillation is not yet triggered. The self-oscillation phenomena confirms the existence of power gain in the nonlinear kinetic inductance superconducting resonator even when the resonator mode is hybridized with the magnon mode. A quantitative analysis of the self-oscillation spectra, including the power, linewidth and frequency period, requires more information and assumption of the intrinsic nonlinearity and damping rate, and is beyond the scope of this work.

In the self-oscillation regime, cascaded wave mixing happens in a much wider frequency range. Figures 4(b) and (c) show the comparison of spectra when a probe tone is applied in the self-oscillation regime with $f_{pump}$=4.318 GHz that is used in Figure 4(a). At low pump power such as $P_{pump}$=-20 dBm where self-oscillation still does not show up, the power spectrum only shows two peaks at $f_{pump}$=4.318 GHz and $f_{probe}$=4.3342 GHz, without any additional peak. As $P_{pump}$ is further increased so that the system enters the self-oscillation regime, besides the periodic self-oscillation peak as marked by aseterisk (*) in Figs. 4(b) and (c), the $f_{probe}$ mode triggers a few additional harmonic peaks, marked by pound (#) in Fig. 4(c). The $f_{probe}$-induced harmonics maintain the same frequency spacing with the closest lower-frequency self-oscillation peaks, meaning that the former is created by mixing with the latter. The robust emergence of these self-oscillatory states above the parametric-amplification regime highlights the rich nonlinear dynamics accessible in superconducting hybrid magnonic resonators.

## Conclusions

In summary, we have demonstrated parametric amplification in a superconducting nanowire resonator mediated by kinetic-inductance nonlinearity. Four-wave mixing processes generated under microwave pumping produce amplification of weak probe signals, yielding gains of up to approximately 7 dB in the magnon-detuned resonator regime. These observations establish the nanowire resonator as an effective kinetic-inductance-based parametric amplifier that does not rely on Josephson junctions. When the resonator is tuned into the strong magnon–photon coupling regime through interaction with a YIG sphere, the nonlinear dynamics become significantly modified by hybridization. The hybrid magnon–photon modes continue to support four-wave mixing and parametric amplification, with gains approaching ~3.4 dB at the anticrossing region. Although the achievable gain is reduced compared with the bare resonator due to the additional magnonic contribution and associated dissipation, the hybridized system provides a unique degree of control over the spectral position of amplification. By tuning the magnon–photon interaction through the applied magnetic field, the operating frequency of the parametric amplification process can be continuously reconfigured without altering the device geometry. Furthermore, we show that the nonlinear response is strongly dependent on the pump power. While moderate pump powers support controlled parametric amplification, increasing the drive strength beyond a threshold results in the emergence of self-sustained oscillatory states characterized by equally spaced sidebands and Hopf-like nonlinear dynamics. These observations reveal the rich interplay between kinetic-inductance nonlinearity, strong magnon–photon coupling, and driven nonequilibrium dynamics. Our results establish hybrid magnonic superconducting resonators as a promising platform for tunable microwave parametric amplification and nonlinear signal processing. More broadly, they demonstrate that magnon–photon hybridization can be employed not only to manipulate coherent microwave excitations but also to engineer and control nonlinear amplification processes, opening new opportunities for reconfigurable microwave and quantum technologies.

## Methods

### Film preparation and device fabrication

Different superconducting films (such as NbN, NbTiN, TiN, or granular Al) have very high upper critical fields and show a uniform Kerr-type nonlinearity without requiring Josephson junctions[8–10]. Hence, we have chosen NbN as our superconducting material, and we deposited it on high-resistivity Si substrates by reactive dc magnetron sputtering from a Nb target. The deposition was carried out at a dc power of approximately 200 W and a working pressure of 3 mTorr with a nitrogen flow rate of 3 sccm. The base pressure prior to deposition was below $2\times10^{-9}$ Torr. During growth, an RF substrate bias of approximately 15 W was applied to improve film quality. Two NbN film thicknesses, 35 nm and 260 nm, were prepared for fabrication of the superconducting resonator and microwave antenna structures, respectively. The superconducting nanowire resonator was defined from the 35-nm-thick NbN film using optical lithography followed by reactive ion etching. The resonator consisted of a 500-nm-wide nanowire incorporating a circular loop designed for strong magnetic coupling to a YIG sphere. To position the magnetic element precisely at the region of maximum microwave magnetic field, cylindrical recesses were fabricated at the center of the loop structure using deep reactive ion etching. The etched cavities had a diameter of 250 μm and a depth of 125 μm, matching the dimensions of the YIG spheres employed in this study. The spheres were fixed within the cavities using a small amount of diluted GE varnish. A second lithography and etching process was performed on the 260-nm-thick NbN film to define a loop-shaped microwave antenna with a diameter of 2.4 mm. The antenna geometry was designed such that the in-plane nanowire resonator could be positioned within the antenna loop, maximizing the overlap between the microwave magnetic fields generated by the two structures. The resonator chip and antenna chip were subsequently assembled using custom copper sample holders, which provided accurate alignment and independent microwave access through SMA connectors for transmission and pump–probe measurements.

### Measurement details

Microwave characterization was performed at a temperature of 1.3 K in a three-axis superconducting magnet cryostat system (American Magnetics Inc.). The transmission characteristics of the device were measured using a Keysight PNA vector network analyzer (VNA). Unless otherwise specified, an intermediate-frequency (IF) bandwidth of 200 Hz was used for all VNA measurements. For nonlinear pump–probe experiments, a continuous-wave pump tone generated by an external microwave synthesizer was combined with the weak probe signal using a power splitter and delivered to the device through the same microwave input line. The resulting mixing products and spectral response were analyzed using a spectrum analyzer. During spectrum-analyzer measurements, the resolution bandwidth was fixed at 1 kHz, while the video bandwidth was maintained at 500 kHz. An external magnetic field was applied perpendicular to the length of the in-plane nanowire resonator, with pump and probe signals fed via antennae as illustrated schematically in Figure 1(a). This field geometry enabled controlled tuning of the YIG ferromagnetic resonance through the superconducting resonator mode, thereby providing access to both the magnon-detuned and strong magnon–photon coupling regimes investigated in this work.

**Acknowledgements**

We sincerely thank Philip Kim and Wolfgang Pfaff for useful discussion and support on the experiment. This work was supported by the U.S. DOE, Office of Science, Basic Energy Sciences, Materials Sciences and Engineering Division under contract No. DE-SC0022060. The research performed in the work, including experimental design, device fabrication and cryogenic microwave characterization, was primarily supported by the U.S. Department of Energy Office of Science, under Contract No. DE-AC02-06CH11357. K.-J.K. is supported by KAIST-funded Global Singularity Research Program for 2021 and the National Research Foundation of Korea (NRF) funded by the Korean Government (MSIP) under grant RS-2023-00275259, RS-2023-00207732. M.S. was supported by the education and training program of the Quantum Information Research Support Center, funded through the National research foundation of Korea (NRF) by the Ministry of Science and ICT (MSIT) of the Korean government under grant No.2021M3H3A103657313.

**Author contributions**

P.K.P. and Y. L. conducted the experiment. M.S, J.Lim, U.W. and Y.L. reviewed and edited the manuscript written by P.K.P. S.K. grew the NbN films, R.D. assisted the resonator fabrications, U.W. assisted the cryogenic measurements, J. Li. assisted the data analysis and useful discussions, C.C and K. J. K. assisted the experiment configuration, V.N., A.H., and Y.L.

supervised the entire project. All authors contributed to the result discussion and manuscript writing.

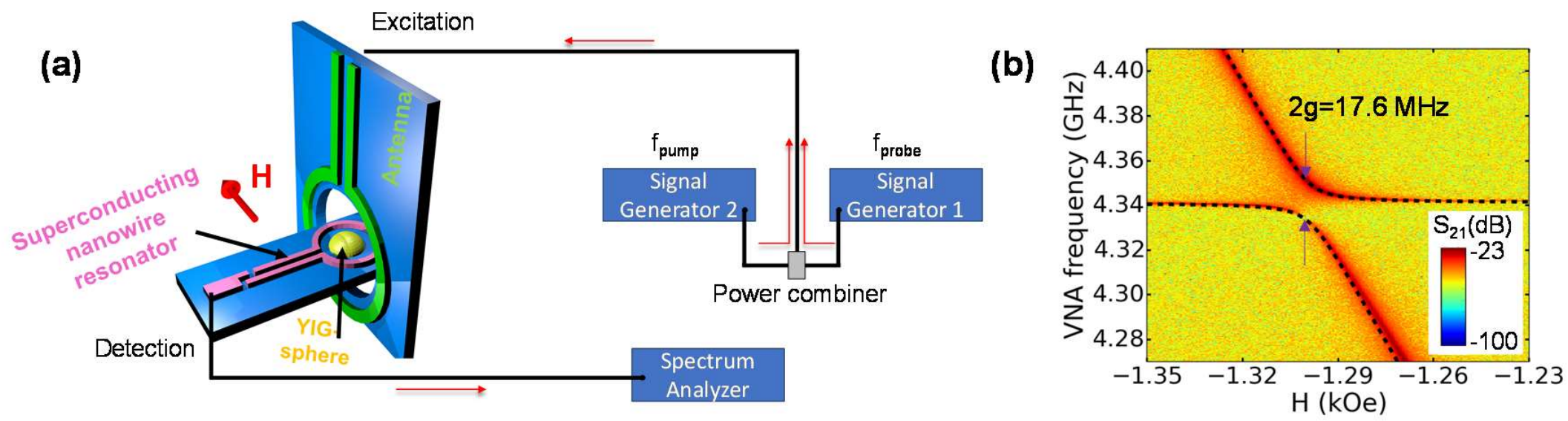

**Figure 1** (a) Schematic of the device structure and the 4-wave mixing (4WM) mechanism consisting of the pump and probe excitation using two signal generators from the antenna port. The magnetic field has been applied at a direction parallel to both the resonator and antenna plane. (b) VNA transmission measurement data between excitation and detection ports of the superconducting hybrid magnonic resonator showing magnon-photon mode anti-crossing.

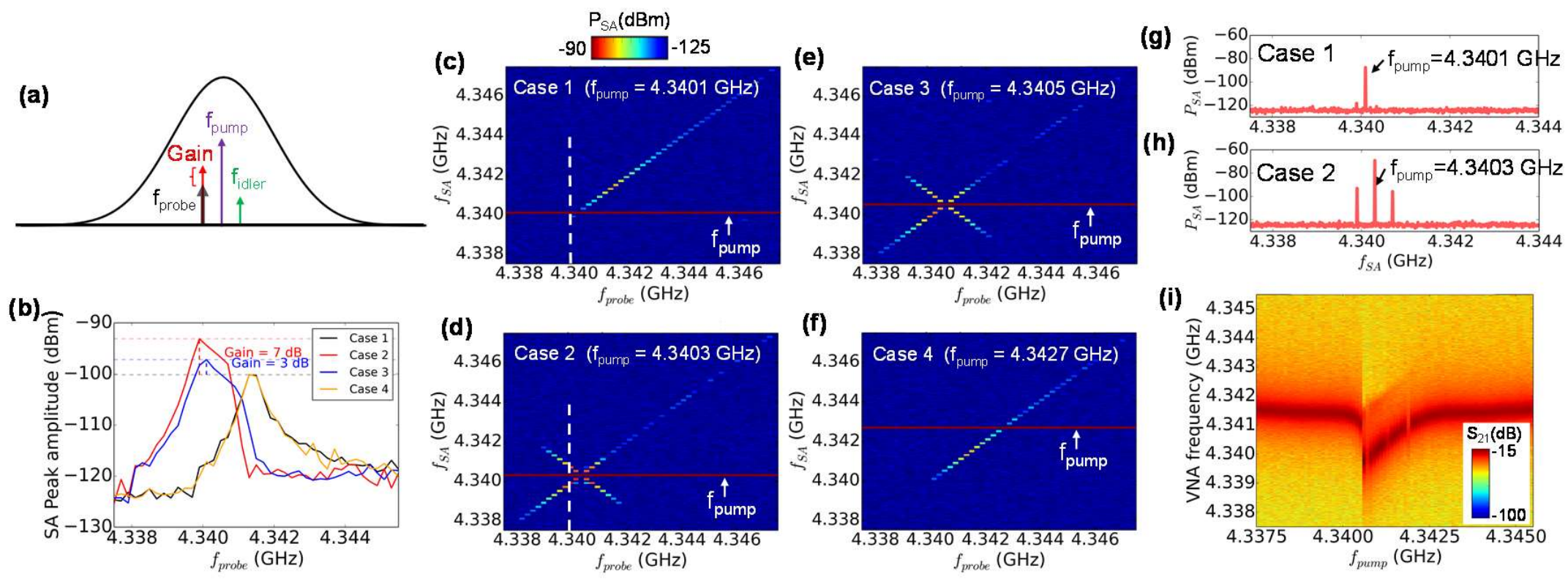

**Figure 2** (a) Illustrates the 4WM mechanism schematic, where the pump transfers its energies to amplify the signal/probe frequency and generates an additional idler frequency at equal spacing. (b) Clearly shows the observed gain due to the parametric amplification phenomenon. (c)-(f) The observed mixing conditions are shown in figures (d),(e), and non-mixing conditions at (c),(f). Figures (g) and (h) show two raw SA data obtained at two specific conditions in white dashed lines in (e) and (c), respectively. The color map for these four 2d plots are shown at the top. (i) Two-tone VNA measurement of the detuned superconducting resonator as a function of pump frequency at H = 0 kOe at $P_{pump}$ = -22 dBm.

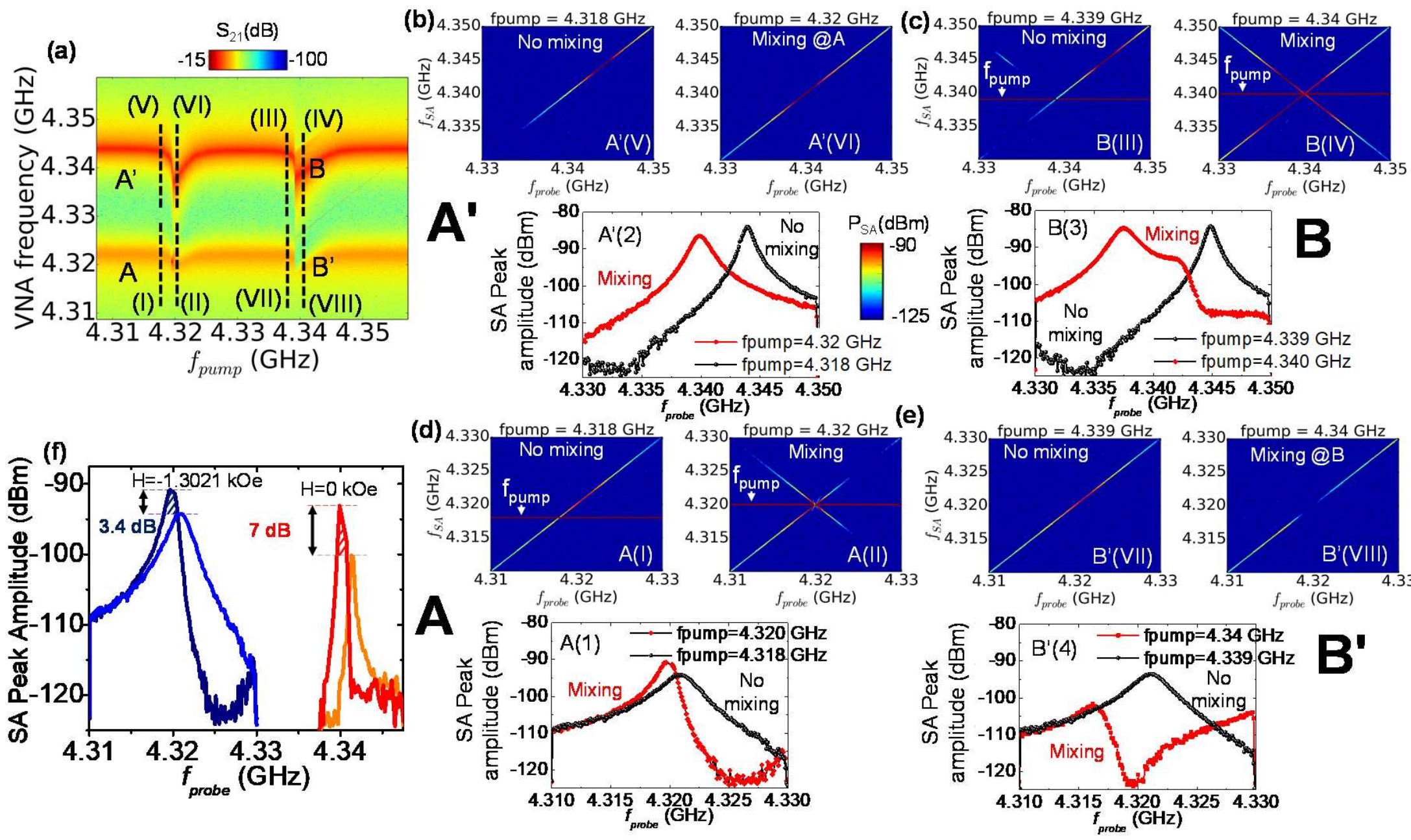

**Figure 3** (a) Two-tone VNA measurement of the superconducting hybrid magnonic resonator as a function of pump frequency at H = -1.3021 kOe which is close to the magnon–photon degeneracy condition. The pump power is fixed as $P_{pump}$ = -22 dBm. The black dashed vertical lines indicates positions where the SA measurements were carried out and shown in (b-e), which consists of four focused regions exhibiting nonlinear frequency pulling, labelled as A, A′, B, and B′. Figures A(1), A'(2), B(3) and B'(4) shows the peak amplitude against $f_{probe}$ for mixing and non-mixing conditions in respective figures. The color map for these eight 2d plots are shown at the centre. (f) Comparison of the peak probe amplitudes measured under non-mixing and mixing conditions for the magnon-detuned resonator (H = 0 kOe) and at the strong-coupling condition (H = -1.3021 kOe). The dashed lines indicate the enhancement of the probe response arising from four-wave-mixing-induced parametric amplification in the two regimes.

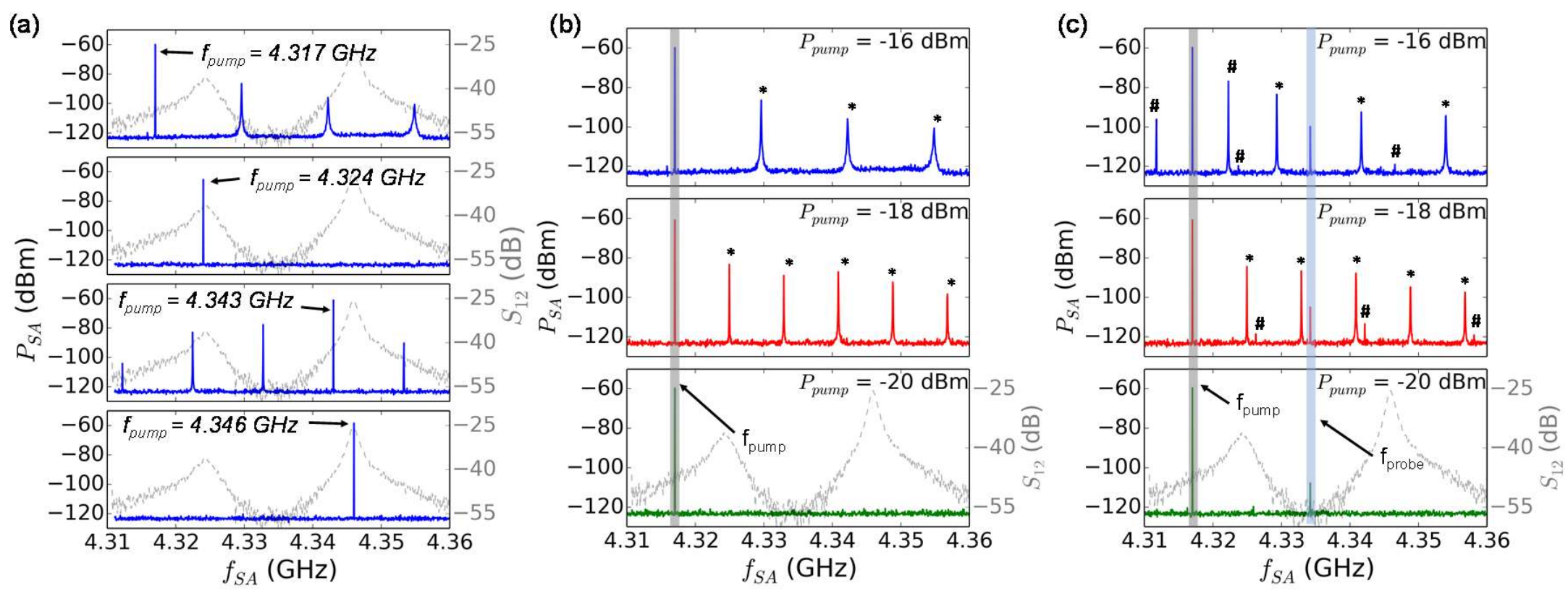

**Figure 4** (a) SA spectra as a function of frequency for different $f_{pump}$ and constant $P_{pump}$ = -16 dBm with probe OFF condition. SA spectra as a function of frequency measured at $f_{pump}$= 4.317 GHz (grey shaded region) at different pump powers ($P_{pump}$), for probe OFF condition in (b) and (c) probe ON at 4.3342 GHz (blue shaded region). and $f_{probe}$= 4.3342 GHz for different pump powers. # denotes additional peaks arising due to the self-oscillation at probe-ON conditions compared to the * peaks at probe-OFF conditions. The VNA measurement has been shown in grey plots to show where we pumped.